\documentclass[twocolumn]{aa}

\usepackage{graphicx}
\usepackage{amsmath}
\usepackage{amsfonts}
\usepackage{amssymb}
\usepackage{gensymb}
\usepackage{textcomp}
\usepackage[varg]{txfonts}
\usepackage[T1]{fontenc}
\usepackage{natbib}
\usepackage{epstopdf}
\usepackage{ulem}
\usepackage{booktabs}
\usepackage{hyperref}
\hypersetup{
    colorlinks,
    citecolor=blue,
    filecolor=blue,
    linkcolor=blue,
    urlcolor=blue,
    menucolor=black}
\newcommand{\RomanNumeralCaps}[1] {\MakeUppercase{\romannumeral #1}}

\makeatletter
\renewcommand*\aa@pageof{, page \thepage{} of \pageref*{LastPage}}
\makeatother

\begin{document} 

\title{On the Effect of Coronal Rain on Decayless Kink Oscillations of Coronal Loops}
\author{Arpit Kumar Shrivastav
      \inst{1,2,3},
      Vaibhav Pant\inst{1}
      \and
      Patrick Antolin \inst{4}
      }

\institute{Aryabhatta Research Institute of Observational Sciences, Nainital, India-263002
    \and 
        Joint Astronomy Programme and Department of Physics, Indian Institute of Science, Bangalore 560012, India
    \and
         Indian Institute of Astrophysics, Bangalore, India-560034
    \and  
        Department of Mathematics, Physics and Electrical Engineering, Northumbria University, Newcastle Upon Tyne, NE1 8ST, UK
         }

   \date{Received **; accepted **}
  \abstract{ Decayless kink oscillations are ubiquitously observed in active region coronal loops with an almost constant amplitude for several cycles. Decayless kink oscillations of coronal loops triggered by coronal rain have been analysed, but the impact of coronal rain formation in an already oscillating loop is unclear. As kink oscillations can help diagnose the local plasma conditions, it is important to understand how these are affected by coronal rain phenomena.  In this study, we present the analysis of an event of coronal rain that occurred on 25 April 2014 and was simultaneously observed by \textit{Slit-Jaw Imager} (SJI) onboard \textit{Interface Region Imaging Spectrograph} (IRIS) and \textit{Atmospheric Imaging Assembly} (AIA) onboard \textit{Solar Dynamic Observatory} (SDO). The oscillation properties of the coronal loop in AIA are investigated before and after the appearance of coronal rain in SJI. We find signatures of decayless oscillations before and after coronal rain at similar positions to those during coronal rain. The individual cases show a greater amplitude and period during coronal rain. The mean period is increased by 1.3 times during coronal rain, while the average amplitude is increased by 2 times during rain, in agreement with the expected density increase from coronal rain. The existence of the oscillations in the same loop at the time of no coronal rain indicates the presence of a footpoint driver. The properties of the observed oscillations during coronal rain can result from the combined contribution of coronal rain and a footpoint driver. The oscillation amplitude associated with coronal rain is approximated to be 0.14 Mm. The properties of decayless oscillations are considerably affected by coronal rain, and without prior knowledge of coronal rain in the loop, a significant discrepancy can arise from coronal seismology with respect to the true values.
  }
  \titlerunning{Effect of Coronal Rain on Decayless Kink Oscillations}
   \authorrunning{Shrivastav et al. 2024}
   \maketitle
   \nolinenumbers
\section{Introduction} \label{sec:intro}
Coronal loops are building blocks of the solar corona, and dense confined plasma gives rise to their brightness (\citealt{Fabio}). The non-thermal velocities, along with up-flows, have been observed near footpoints of the active region coronal loops, suggesting footpoint heating (\citealt{Doscheck2007}, \citealt{hara2008}). Thermal conduction distributes the footpoint heating to the chromosphere, which heats the plasma, leading to chromospheric evaporation. The density increases in the corona, which ends up being thermally imbalanced, with radiative cooling dominating the heating and driving the loop into a state of thermal non-equilibrium (TNE) cycle, which is a global state of a coronal loop (for a detailed review of the TNE cycle, please refer \citealt{Antolin2020} and \citealt{Antolin&Froment2022}). During cooling, thermal instability (TI) can set in at a specific density and temperature value. Due to this process,  the loop apex cools down catastrophically to chromospheric and transition region temperatures, forming plasma clumps within a few minutes. These clumps fall along the loop, guided by the magnetic field lines, and are known as coronal rain. The first reports of coronal rain date back to the 1970s (\citealt{kawaguchi1970}, \citealt{Leroy1972}), although \cite{1938McMath} provide the description of a phenomenon that strongly resembles coronal rain. When seen on the limb, the coronal rain clumps appear as bright features in the chromospheric lines such as H$\alpha$ (\citealt{De_Groof}, \citealt{De_Groof2005}), Ca II H (\citealt{Antolin2010}, \citealt{antolin&verwitche2011}) or transition region lines such as He II, Si IV(\citealt{De_Groof2005}, \citealt{Antolin2015}). They appear as dark structures when observed on-disk (\citealt{Antoline_ondisk2012}). They fall towards the solar surface with acceleration less than free-fall gravitation acceleration, and numerical simulations indicated increased pressure gradient downstream of the rain to be one of the reasons (\citealt{Muller2004}, \citealt{Antolin2010}, \citealt{Oliver2014}). Earlier, coronal rain was considered a sporadic phenomenon with frequency in days (\citealt{Schrijver2001}), but recent estimates of occurrence times using high-resolution observations with the \textit{Swedish 1-m Solar Telescope} showed that coronal rain is a common phenomenon of active regions (\citealt{Antolin&van2012}). More recently, a strong link between coronal rain and the phenomenon of long-period intensity pulsations has been established (\citealt{Auchere2018}). The latter corresponds to highly periodic EUV pulsations observed along coherent structures (such as loops) lasting several days (\citealt{Auchere2014}, \citealt{Froment2015}), now understood as the manifestation of TNE cycles in loops (\citealt{Froment2017,Froment2018}, \citealt{Pelouze2022}). Coronal rain can, therefore, be recurrent in the same loop bundle with a timescale of hours. The density of coronal rain clumps varies between $10^{10} - 10^{12} \text{cm}^{-3}$ \citep{Antolin2015} and, on average, have widths of hundreds of km and lengths reaching up to a few Mm. The formation of coronal rain carries important observable features in the extreme ultraviolet (EUV). As the rain forms, due to flux freezing, magnetically enhanced strands strongly emitting in the UV and EUV can form, linked to the condensation-corona transition region (CCTR) at the boundaries of the rain. Furthermore, significant compression can occur downstream from the rain, leading to strongly variable plume-like structures \citep{2022Antolin-martinez-sahin}.

Coronal loop oscillations are powerful tools for diagnosing coronal parameters through magnetohydrodynamic (MHD) seismology. Transverse oscillations in coronal loops have been interpreted as fast MHD kink waves (\citealt{Edwin&Roberts}, \citealt{Roberts&Edwin&Benz1984}). \textit{Transition Region and Coronal Explorer} (TRACE) observed the earliest transverse oscillations of coronal loops in  EUV wavelengths (\citealt{Aschwanden1999}, \citealt{Schrijver1999}, \citealt{Nakariakov1999}). Coronal rain, prominences, and similar chromospheric flows occurring in the corona have also been used as tracers of transverse MHD waves (\citealt{Okamoto2007}, \citealt{OfmanandWang2008}, \citealt{antolin&verwitche2011}, \citealt{Kohutova&Verwichte2016}, \citealt{Verwitcheantoline2017}, \citealt{verwichte2017}). Furthermore, the ultra-long period oscillations (10-30 hr), interpreted in terms of MHD waves, have been observed in filaments \citep{2004Foullon, 2009Foullon}.  The first identification of transverse oscillations in coronal rain was performed by \cite{antolin&verwitche2011} using the Hinode/ Solar Optical Telescope. These oscillations have amplitudes below 500 km and periods in the range of 100 to 200 s. The oscillations were interpreted as the first harmonic of the standing kink MHD mode because the amplitude of oscillations was maximum at midway between the loop apex and footpoint. Coronal loops have been seen to exhibit persistent low-amplitude oscillations that show no apparent decay over multiple periods, known as decayless kink oscillations. They are frequent features in the solar corona \citep{2015Anfinogentov, Gao, 2024Shrivastav}. The study of \cite{Kohutova&Verwichte2016} revealed the small amplitude, persistent transverse oscillations of coronal rain with an average period of 3.4 minutes and amplitude lying between 0.2-0.4 Mm.

\cite{verwichte2017} reported the first observation of excitation of kink oscillations by coronal rain. The study indicated that the oscillations become apparent as condensation forms due to the significant coronal rain mass. \cite{Kohutava&Verwitche2017} performed a 2.5 dimensional numerical MHD simulation of a coronal loop by introducing low temperature, high-density condensed mass at the loop top. The loop was displaced downwards, setting up the transverse oscillation. The study showed that the amplitude of the excited transverse oscillations increases with rain mass compared to the total loop mass. This study also revealed that as the rain blobs fall towards the loop leg, the period of kink oscillations decreases due to the temporally varying average density of the loop. The generation of transverse MHD waves by coronal rain was first studied analytically by \cite{Verwitcheantoline2017}. They presented a mechanical model of coronal rain in which the role of the ponderomotive force arising from the transverse oscillations on the kinematics of rain blobs was investigated. It was found that the low amplitudes of transverse oscillations previously seen in rain events are not enough to account for the reduced downward acceleration. The model indicated that the amplitude of excited transverse oscillations due to the rain would depend on various parameters and be affected by the time scales of rain formation. 
\begin{figure*}[!ht]
    \centering
    \centerline{
               \includegraphics[trim= 0 0.85cm 0 0 ,width=1.2\textwidth,clip]{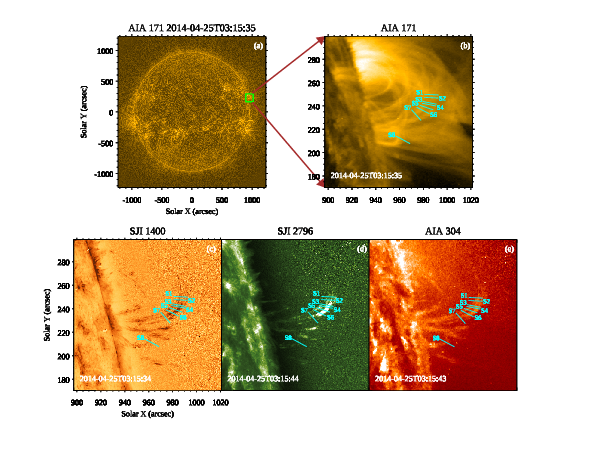}
               
              }
    \caption{(a) Full disk intensity image of the Sun in AIA 171 processed using MGN technique. The small green box highlights the event location and FOV of SJI. (b) Zoomed view of the green box in AIA 171 \AA\ with the loop system.
    (c)-(e) SJI 1400, 2796, and AIA 304 (radial gradient filtered) images approximately at the same time. The presence of coronal rain in the loops can be seen. The cyan lines mark the position of artificial slits chosen for generating space-time maps. Slit 8 overlaps the position of a smaller loop. An animation associated with the figure is \href{https://drive.google.com/file/d/16_LjLjiuD4RcSO0h_I2FlzGvvDOA3qQ-/view?usp=sharing}{available}. }
    \label{fig:fig1}
\end{figure*}
Although kink oscillations and coronal rain are common phenomena in active region loops, the cause-and-effect relationship is unclear. For instance, although coronal rain can excite transverse MHD oscillations, it is not clear how common this process is in the solar corona. We present the observations of transverse oscillations accompanied by coronal rain and simultaneously observed by the \textit{Slit-Jaw Imager} (SJI) and \textit{Atmospheric Imaging Assembly} (AIA). We applied the motion magnification technique to magnify transverse oscillations in the AIA channels, which allowed us to detect and compare the properties of oscillations before, during, and after coronal rain. 

The paper is arranged as follows: Section 2 covers the details of observation and data. In Section 3, we present the methodology, individual cases of oscillations before and after coronal rain, and a comparison of the average properties of oscillations. In Section 4, we summarised the work and discussed the possible scenarios leading to the observed results.

\section{Observational data} \label{sec:obs}

 \begin{figure*}[!ht]
    \centering
    \centerline{
               \includegraphics[trim= 0.05cm 1.2cm 0 1.5cm ,width=0.9\textwidth,clip]{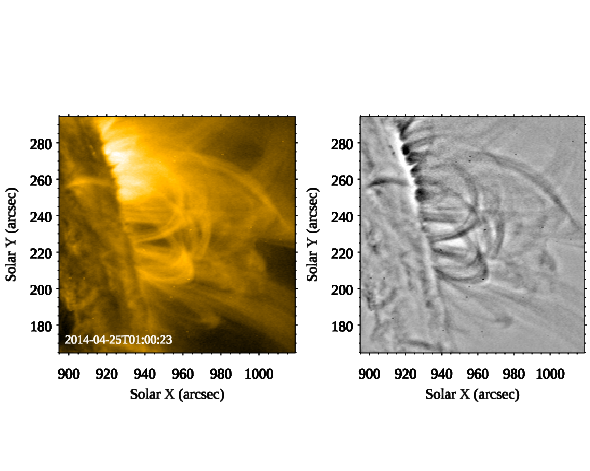}
               
              }
    \caption{The left panel shows the AIA 171 intensity image of the small green box shown in Figure \ref{fig:fig1}. The fuzziness of emission is clearly visible. The right panel shows the unsharp masked inverted image of the same FOV in which loop edges are enhanced. } 
    \label{fig:fig2}
\end{figure*}

We studied coronal rain observed by the \textit{Slit-Jaw Imager}(SJI) onboard Interface Region Imaging Spectrograph \citep[IRIS;][]{DePontieu2014} at the north-east limb on April 25, 2014. IRIS started observing the region around 00:39 UT and observed the event in two passbands, 1400 (far-UV) and 2796 (near-UV). The emission in the SJI 1400 passband is contributed by Si \RomanNumeralCaps{4} emission lines that form at the transition region temperature of about 63000 K, whereas the SJI 2796 passband is dominated by Mg \RomanNumeralCaps{2} K line core that forms at the chromospheric temperature of about 10000 K. The dataset has a pixel scale of 0.166\arcsec\ $\text{pixel}^{-1}$ and a cadence of 18.7 s. IRIS observations are centered around [x,y]=[957\arcsec,229\arcsec]. 

The event is also observed by \textit{Atmospheric Imaging  Assembly}  \citep[AIA;][]{Lemen2012} on board Solar Dynamic Observatory \citep[SDO;][]{SDO2012}. AIA takes full-disk, nearly simultaneous observation of the Sun in seven EUV narrow-band filters. AIA has a pixel-scale of 0.6\arcsec\ $\text{pixel}^{-1}$ and a cadence of 12 s. AIA data is de-rotated, co-aligned, and brought to the same pixel scale using aia\_prep.pro. Coronal rain condensations are spotted in AIA 304, dominated by He \RomanNumeralCaps{2} (304 \AA) emission at approximately equal to 50,000 K. The loop system was best visible in the AIA 171 passband (dominated by Fe \RomanNumeralCaps{9} 171.09 Å emission at 0.7 MK) compared to other AIA channels. Thus, we use 304 and 171 passband images to analyse this event.

 Figure \ref{fig:fig1}(a) shows the AIA 171 intensity image processed using Multi-Gaussian Normalization (MGN) technique (\citealt{Morgan&miller2014}) at 03:15 UT. The green rectangle represents the field of view (FOV) observed by SJI, and Figure \ref{fig:fig1}(b) shows a zoomed view of the same FOV with the loop system observed in AIA 171. Panels (c)-(e) in Figure \ref{fig:fig1}  show the coronal rain in SJI 1400, 2796 and AIA 304, respectively. A radial gradient filter similar to a normalizing radial graded filter \citep{Morgan&habbal} is applied to SJI 2796, 1400, and AIA 304 intensity images to enhance off-limb intensity. AIA 171  intensity images look fuzzy due to the diffused background and foreground emission (\citealt{2003Delzanna}, \citealt{Viall})  produced in an active-region coronal loop system with a peak temperature of about  1.5 MK (\citealt{Subramanian}, \citealt{Brooks}). Although it is difficult to segregate adjacent loops and their strands, we attempt to separate the loops with diffused emission using unsharp masking to identify oscillating strands, as shown in Figure \ref{fig:fig2}(b). The images are convolved with a Gaussian kernel to remove the noise and then subtracted from the original to identify and enhance edges. 
 AIA and SJI channels are aligned by matching the off-limb features in the near-simultaneous images of SJI 1400 and AIA 304. 

Additionally, we utilize data from the Extreme Ultraviolet Imager \citep[EUVI;][]{2004Wuelser_EUVI} onboard the Solar Terrestrial Relations Observatory  \citep[STEREO;][]{2008Kaiser_stereo}. The EUVI captures full disk images of the Sun in four passbands, with a pixel scale of 1.6 \arcsec\ $\text{pixel}^{-1}$. For our analysis, we rely on the 171 \AA\ images to estimate the three-dimensional geometry of the loop. During the observation period, EUVI 171 provided only two images.

\section{Analysis and results} \label{sec:analysis} 

\begin{figure*}[!ht]
    \centering
    \centerline{
    
               \includegraphics[trim=0.2cm 4.8cm 0cm 0.5cm ,width=\textwidth,clip]{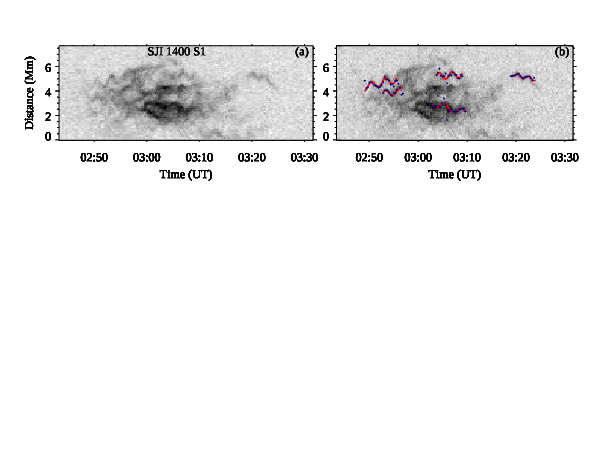}
               
              }
    \caption{(a) x-t maps of Slit 1 near the apex of the loops in SJI 1400 shown in Figure \ref{fig:fig1}. The negative intensity is shown in each map. These x-t maps show that the different rain clumps oscillate in phase at varying positions of the slits. The decay-less nature of transverse oscillations is also visible in the x-t maps. (b) The plus signs are centres of the threads over-plotted with the obtained sinusoidal fits, represented by red curves. The oscillation properties are derived using sinusoidal fits.}
    \label{fig:fig3}
\end{figure*}

\subsection{Space-Time (x-t) maps}
We placed eight artificial slits over two coronal loops at different positions to capture oscillations before, during, and after coronal rain. Figure \ref{fig:fig1}(b)-(e)  shows the position of slits as cyan lines orthogonal to the loop in different passbands. Slits 1 to 7 are placed on the larger loop, while the second loop hosting slit 8 is smaller.  These artificial slits are five pixels wide, and intensity is averaged across the artificial slit to increase the signal-to-noise ratio. Figure \ref{fig:fig3}(a) shows the x-t maps for slit 1 in SJI 1400, and it is evident that different blobs are subjected to transverse oscillations. These x-t maps also indicate that these oscillations are decayless, and we did not observe any eruption or flare near the loop system for about 3 hours before the start of IRIS observation. It is important to highlight that the rain blobs observed between 02:50-02:56 UT in Figure \ref{fig:fig3}(a) are oscillating in phase, indicating the potential existence of standing kink oscillations.

\begin{figure*}[!ht]
    \centering
    \centerline{
               \includegraphics[trim=0cm 2.45cm 0cm 1cm ,width=1.2\textwidth,clip]{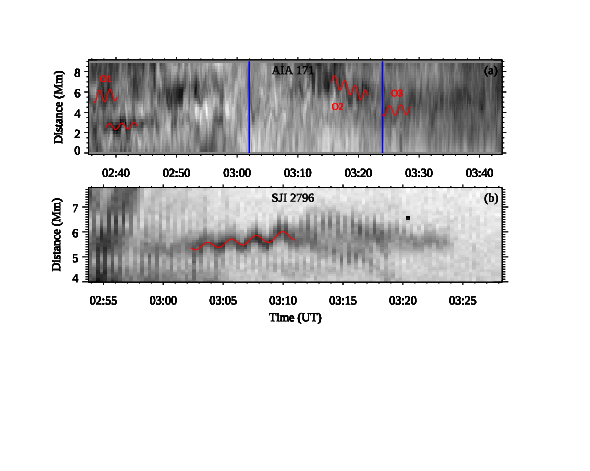}
               
              }

    \caption{(a) AIA 171 x-t map of Slit 8 for the time before, during, and after coronal rain. The blue vertical lines indicate the time interval when coronal rain was present in the SJI 2796 passband. The x-t maps for AIA channels are produced after motion magnification. These xt-maps are inverted in intensity. (b) the x-t map of Slit 8 in SJI 2796. The amplification in amplitude is clearly visible, and oscillation is fitted with the red curve. }
    \label{fig:fig4}
\end{figure*}

To determine oscillation parameters, we estimate the oscillating thread centre by fitting a Gaussian function over the distance along the slit at each time step. The centres of best-fit Gaussian are overplotted with a ``plus" sign over time distance maps (see Figure \ref{fig:fig3}b). We fitted the oscillations using the following relation. 
\begin{equation}
    A(t) = A_{0}+A_{1} exp{\left(\frac{t}{\tau}\right)} \sin \left(\frac{2\pi t}{P+kt}+\phi \right)+A_{2}t.
\end{equation}
Where $A_{0}$ is the mean position, $A_{1}$ is displacement amplitude, $P$ is time period, $\phi$ is the phase, $A_{2}$ is a constant parameter for linear trend, $k$ is a coefficient for the linear increase or decrease in the period of oscillations and $\tau$ provides damping or growth time scale. We use the Levenberg-Marquardt least-squares minimization to obtain fitting parameters. The obtained fits for slit 1 are plotted with the centres in Figure \ref{fig:fig3}(b). We employ the width of the fitted Gaussian as uncertainty on the position of the loop. The amplitude and period are also supplemented with errors obtained after sinusoidal fitting. The parameters of the fitted oscillations are provided in Table \ref{Tabel1}.

\subsection{Motion magnification}
We find oscillations with an average amplitude less than 1 Mm in SJI 1400 and 2796 x-t maps. It was challenging to find oscillations at the same spatial positions in AIA channels as these amplitudes are comparable to the pixel resolution of AIA. \cite{2015Anfinogentov} showed that low amplitude decayless oscillations are common features in solar corona with average displacement amplitude less than a pixel size of AIA. To enhance the transverse motions in AIA observations, we use the motion magnification (\citealt{2016SoPh..291.3251A}) technique, which magnifies transverse oscillation amplitudes while preserving the oscillation periods. This technique was previously used for the identification of second harmonics of decayless kink oscillations (\citealt{Duckenfield}), coronal seismology of quiet Sun using decayless kink oscillations (\citealt{AnfinogentovNakariakov}), investigation of transverse oscillations linked to flares (\citealt{Mandal}) and identification of decayless oscillations in coronal bright points (\citealt{Gao}). In this work, we use a magnification factor of 7 as it is found optimal to identify the oscillations.

\subsection{Individual cases of oscillations before and after coronal rain} \label{bef_aft_rain}

\begin{figure*}[!ht]
    \centering
    \centerline{
               \includegraphics[trim=0cm 2.45cm 0cm 1.1cm ,width=1.1\textwidth,clip]{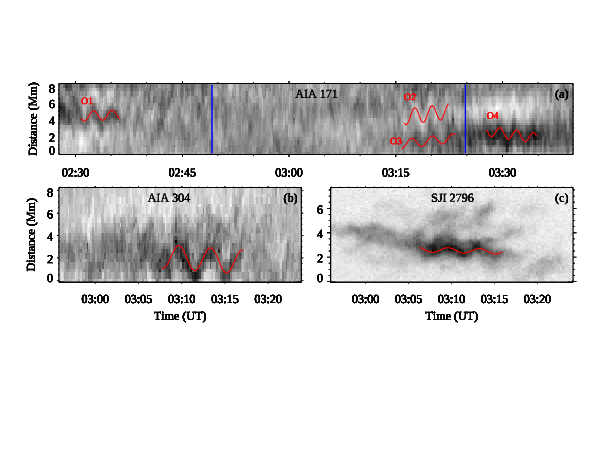}
               
              }
    \caption{(a) AIA 171 x-t map of Slit 3 for the time during and after coronal rain. The two vertical blue lines represent the time span from the first appearance of coronal rain until it reaches the footpoint of the loop in the SJI 2796 channel.  (b)-(c) The oscillation captured in SJI 2796 and AIA 304 for the same slit when coronal rain appeared. The x-t maps for AIA channels are produced after motion magnification. All xt-maps are inverted in intensity. }
    \label{fig:fig5}
\end{figure*}

\begin{figure*}[!ht]
\centering
\includegraphics[width=0.9\textwidth,clip,trim=0cm 4.5cm 0cm 0cm]{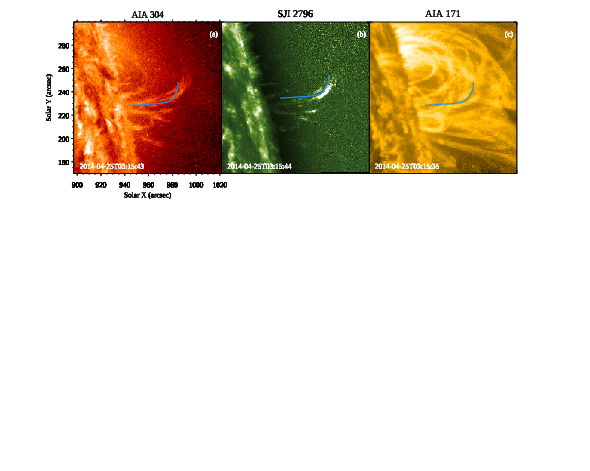}
\includegraphics[width=0.9\textwidth,clip,trim=0cm 0cm 0cm 0.5cm]{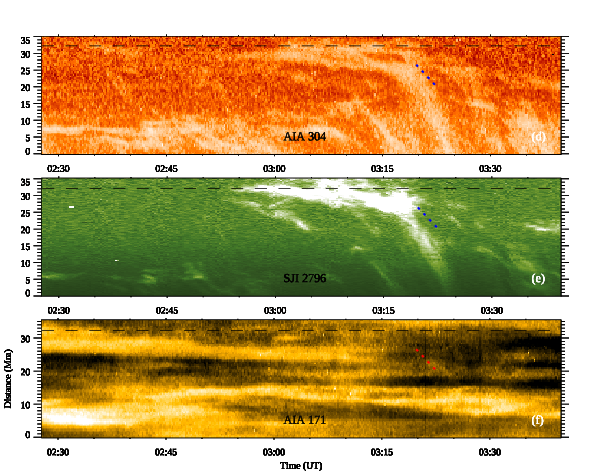}
\caption{(a)-(c) depict the location of the curved slit, traced along the rain blobs in AIA 304, SJI 2796, and along the loop in AIA 171, aiming to detect the presence of rain blobs. The slit's placement is determined based on the trajectory of rain blobs observed in SJI 2796 images. (d)-(f) illustrate the x-t maps associated with the curved slit in AIA 304, SJI 2796, and AIA 171. The black line in the x-t maps represents the estimated position of slit 3 on the curved slit. Around 03:20 UT, the descending rain in EUV absorption is traced by the red dashed line in the AIA 171 x-t map. The line is displayed with an offset to clearly identify the rain features in EUV absorption. The blue dashed line in AIA 304 and SJI 2796 corresponds to the red dashed line in AIA 171.  }
\label{fig:slit_along_rain}
\end{figure*}

\begin{figure*}[!ht]
\centering
\includegraphics[width=\textwidth,clip,trim=0.7cm 4.7cm 1cm 0.6cm]{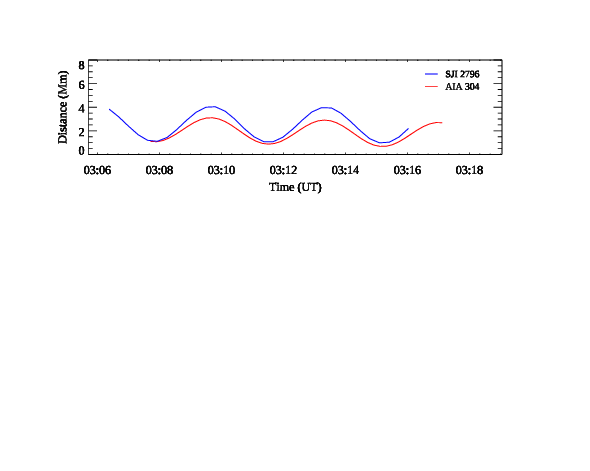}
\caption{The curves representing the oscillations observed in slit 3 during rain are shown for both SJI 2796 and AIA 304. The SJI 2796 oscillation amplitude is scaled by a factor of 7 to facilitate a more meaningful comparison. The phase difference between these oscillations is negligible.  }
\label{fig:phase_lag_304_2796}
\end{figure*}

\begin{figure*}[!ht]
\centering
\includegraphics[width=\textwidth,clip,trim=0cm 4.6cm 0cm 0.9cm]{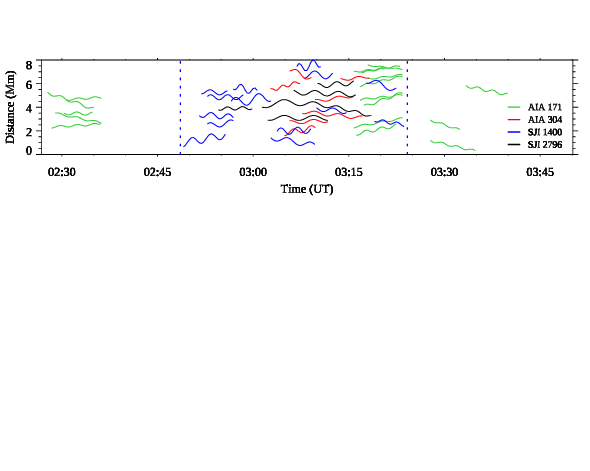}

\caption{All identified oscillations across various channels are plotted on a single timeline with slit distance in the y-axis. The oscillations from different slits are randomly distributed along the slit. The blue dashed lines mark the time interval during which coronal rain is observed in the SJI channels.}
\label{fig:all_oscillations_timeline}
\end{figure*}

\subsubsection{Oscillations in the smaller loop} \label{oscl_small_loop}

 We use the motion magnification algorithm to capture oscillations before and after coronal rain. Figure \ref{fig:fig4}(a) shows x-t maps, generated after motion magnification, corresponding to slit 8 on the smaller loop in AIA 171. The x-t maps are presented in inverted intensity, so dark features correspond to emission. Two blue vertical lines in Figure \ref{fig:fig4}(a) correspond to the time of the start and end of the coronal rain as observed from SJI 2796. The SJI 2796 x-t map shows the signature of oscillations between 4 and 6 Mm. The oscillation observed in slit 8 in Figure \ref{fig:fig4}(b) shows amplitude amplification in time.  We fitted this growing amplitude oscillation using equation 1 in \cite{2012Wang},  which is overplotted in red. The oscillation amplitude and period are computed as 0.11$\pm$0.01~Mm and 2.1$\pm$0.01~minutes, respectively. The growth time scale, $t_{g}$, for oscillation, is 695s. The increase in the amplitude is estimated from ratio $A_{1,t_1}/A_{1,t_0} = e^{(t_1-t_0)/t_{g}} = $2.1, where $t_0$ and $t_1$ are start and end time of observed oscillation.  If we trace back to the similar location in the AIA 171 passband in Figure \ref{fig:fig4}(a), we see an oscillation, O1, between 02:35-02:41 UT with an amplitude of 0.09$\pm$0.01 ~Mm and a period of 1.7$\pm$0.04 ~minutes. We confirm the presence of oscillation signatures around 03:12 UT ({\it, i.e.,} during the coronal rain) in the AIA 304 channel between 4 to 6 Mm. The oscillating blob is not fitted for AIA 304 because it does not have a clear contrast relative to the background.
During coronal rain, we detected the oscillation labelled as O2 in emission from 5 to 7 Mm in the AIA 171 xt map. This can be a possible signature of coronal rain oscillation in CCTR emission.

The coronal rain is observed in SJI 2796 between 03:02 and 03:24 UT. Just after 03:24 UT, we see an oscillation, O3, at about 4 Mm in AIA 171. The estimated amplitude and period are  0.07$\pm$0.01 ~Mm and  1.9$\pm$0.02 ~minutes, respectively. We do not see a clear signature of oscillations after 03:30 UT in AIA 171. The diffused background and foreground and overlapping loops can make it challenging to capture oscillations after motion magnification. The decrease or increase in emission in a particular passband due to the thermal evolution of the loop top can also be the cause of the fading nature of these oscillations. The thermal evolution for the loop top is not discussed for the smaller loop, as the emission in cooler channels was affected by coronal rain in a foreground loop. 

\subsubsection{Oscillations in the bigger loop} \label{big_loop}

Figure \ref{fig:slit_along_rain}(a)-(c) displays the location of the curved slit in AIA 304, SJI 2796, and AIA 171 intensity images taken along the larger loop. Panels (d)-(f) in Figure \ref{fig:slit_along_rain} depict the corresponding x-t maps for the curved slit, with the black dashed line indicating the location of slit 3 on the curved slit. Rain blob signatures are observed near the loop top in AIA 304 and SJI 2796 x-t maps after 02:55 UT. These maps reveal the position of rain blobs along the loop, reaching the bottom of the loop around 03:25 UT. Notably, there is no evidence of rain in the loop between 02:27 and 02:55 UT, as observed in AIA 304 and SJI 2796 x-t maps. This suggests that rain occurred only between 02:55 and 03:25 in the bigger loop. AIA 171 x-t map shows the coronal rain signature in EUV absorption around 03:20 UT, indicated by the dashed red line (see Figure \ref{fig:slit_along_rain}(f)). The line is shifted in the time axis to view the rain in absorption clearly. The corresponding positions of the red line in AIA 304 and SJI 2796 are indicated by blue lines.

Figure \ref{fig:fig5}(a)-(c) shows the x-t maps for slit 3 on the bigger loop in which the x-t maps of AIA 171 and 304 are made after motion magnification. AIA 171 x-t map of slit 3 shows the detected oscillation, O1, around 02:30 UT, while the coronal rain started after 02:55 UT. The oscillation has amplitude and period of 0.08$\pm$0.02 Mm and 2.6$\pm$0.02 minutes, respectively. The oscillation captured by AIA 304 has amplitude and period of 0.16$\pm$0.01 ~Mm and 3.8$\pm$0.09 ~minutes, respectively, though it is 0.22$\pm$0.04 ~Mm and 3.9$\pm$0.03 ~minutes for oscillation captured by SJI 2796. However, when we plot the oscillations detected in slit 3 using SJI 2796 and AIA 304 channels on the same timeline and position along the slit, we find that they line up and have similar phases (see Figure \ref{fig:phase_lag_304_2796}). Furthermore, their mean positions along the slit are close to each other, suggesting that these oscillations are essentially identical.  The oscillations during coronal rain are seen in AIA 304, and SJI 2796 persisted till 03:17 UT (Figure \ref{fig:fig5}(b)-(c)). We found evidence of oscillations at similar positions in AIA 171 x-t maps after coronal rain in different slits. Figure \ref{fig:fig5}(a) shows the oscillation, O4, in the AIA 171 x-t map after 03:25 UT at a spatial position close to that detected in SJI 2796 and AIA 304 x-t maps. The oscillation detected in AIA 171 after coronal rain has an amplitude of 0.09$\pm$0.01 ~Mm and a period of 2.4$\pm$0.02 ~minutes.

\begin{figure}[h]
    \centering
    \includegraphics[trim=0.41cm 4.1cm 3.4cm 0.5cm,width=0.47\textwidth,clip]{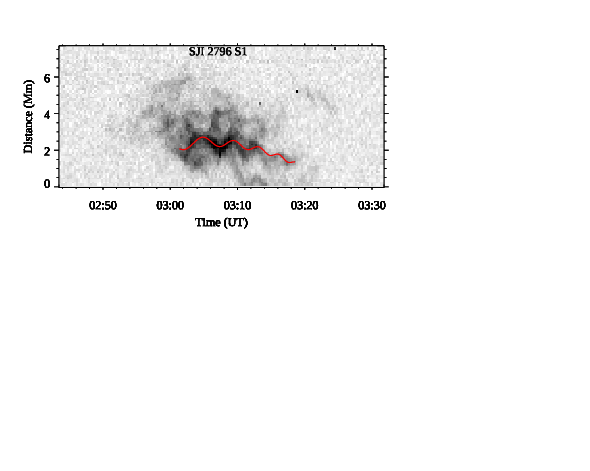}     
    \caption{SJI 2796 x-t map of slit 1. The amplitude and period of oscillation captured in the x-t map decay with time.}
    \label{fig:decay}
\end{figure}

\begin{figure*}[]
    \centering
    \centerline{
               \includegraphics[trim=0cm 0cm 0cm 0cm ,width=\textwidth,clip]{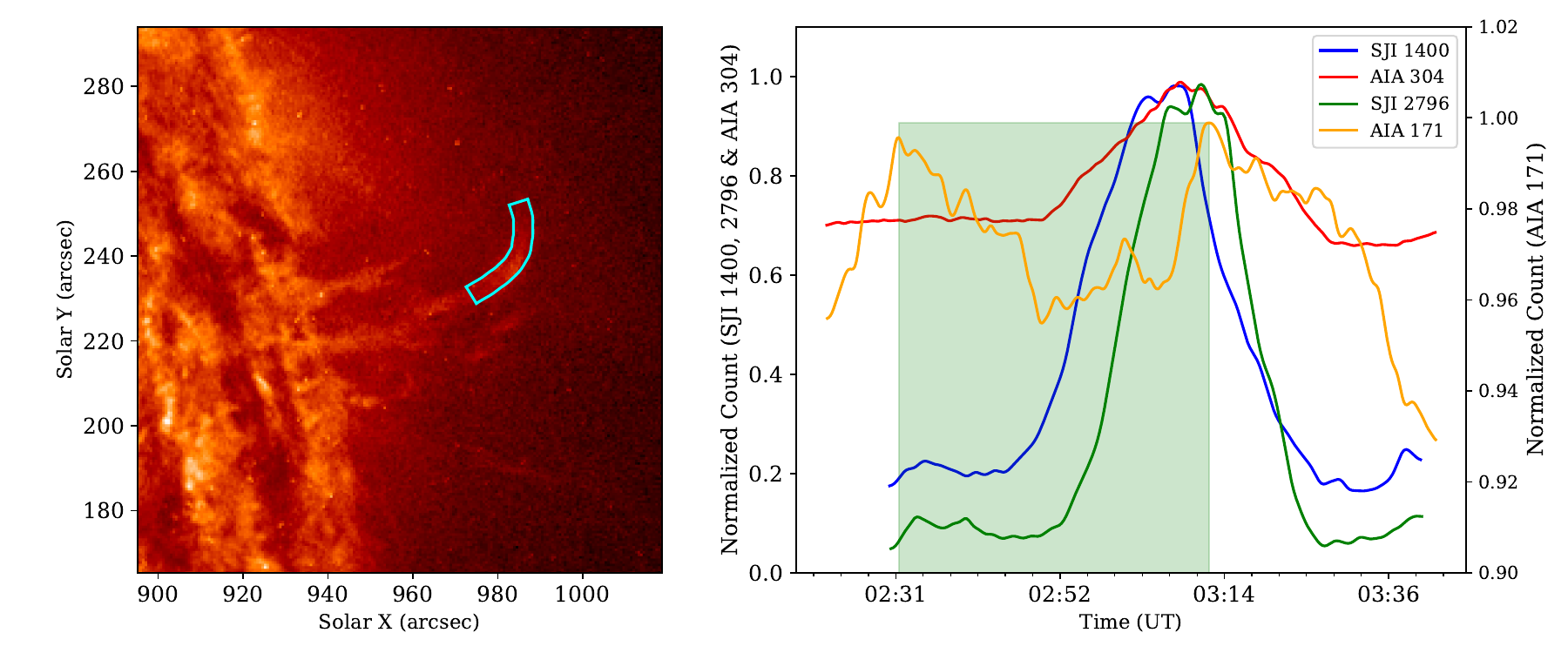}
               
              }
    \caption{The left panel shows the cyan box enclosing the loop top at which slit 3 was placed. The right panel shows the temporal evolution of the intensity summed over the cyan box in four passbands. The green box shows the time interval between the maximum of the counts observed in AIA 171.}
    \label{fig:fig6}
\end{figure*}
 
From 03:15 to 03:25 UT, we detect oscillations, O2 in absorption, and O3 in emission in AIA 171 x-t maps (Figure \ref{fig:fig4}(a)). These oscillations could be attributed to EUV absorption and CCTR emission originating from rain blobs, as we also observe rain oscillations during this time interval in the SJI 1400 channel with similar amplitude and period (see Figure \ref{fig:all_oscillations_timeline}).

Figure \ref{fig:all_oscillations_timeline} presents a compilation of all detected oscillations in SJI and AIA channels within slits 1 to 7, focusing on the bigger loop. The oscillations are arbitrarily shifted along the slit distance for clarity. Within this timeline, two blue dashed lines delineate the time frame during which coronal rain is visible in SJI images. Notably, oscillations detected across various slits exhibit in-phase behaviour before, during, and after the occurrence of coronal rain, indicating their standing nature. Furthermore, no evidence of propagating waves is discerned before or after the coronal rain in AIA 171 $x-t$ maps.

We find several oscillations before and after coronal rain to be decayless. However, as discussed in section \ref{oscl_small_loop}, we observe a case of amplification during coronal rain. Apart from that, we also observed an oscillation with decreasing period and displacement amplitude during coronal rain in slit 1 of the x-t map of SJI 2796 (Figure \ref{fig:decay}). A similar nature of oscillations during coronal rain has been observed previously by \citealt{verwichte2017}.

The individual cases shown here indicate that the amplitude and period are lower before coronal rain, as seen in AIA 171 x-t maps. It is higher during coronal rain, as observed in AIA 304 and SJI channels. During coronal rain, we observe a case of damping and also a case of amplification. We also find lower amplitude and period after coronal rain in AIA 171.

\subsection{Effect of the thermal evolution of the bigger loop on the appearance of oscillations}
AIA 171 x-t map in Figure \ref{fig:fig5}(a) shows that oscillations are not present between 02:38-03:15 UT, while the oscillations are visible between 03:00-03:17 UT in AIA 304 and SJI 2796 (Figure \ref{fig:fig5}(b)-(c)). To investigate the impact of the thermal evolution of the loop on the appearance of oscillations in x-t maps, we placed a curved box surrounding the loop top on the bigger loop (left panel of Figure \ref{fig:fig6}). The average normalized counts in the box are plotted with time for different channels in the right panel. The emission is significant in AIA 304 and SJI 2796 channels when oscillations are visible in the x-t maps for both. The green box shows the time between two maximums in AIA 171 normalized counts. The AIA 171 emission decreases after 02:32 UT, which can be a possible signature of cooling. The AIA 171 emission is lower between 02:32 and 03:14 UT, which is a similar range when oscillations are not present in the AIA 171 x-t map. This suggests that the thermal evolution of the loop can also affect the appearance of oscillations.

\begin{figure}[!ht]
    \centering
    \centerline{
               \includegraphics[trim=0cm 0.3cm 0cm 0.3cm ,width=0.5\textwidth,clip]{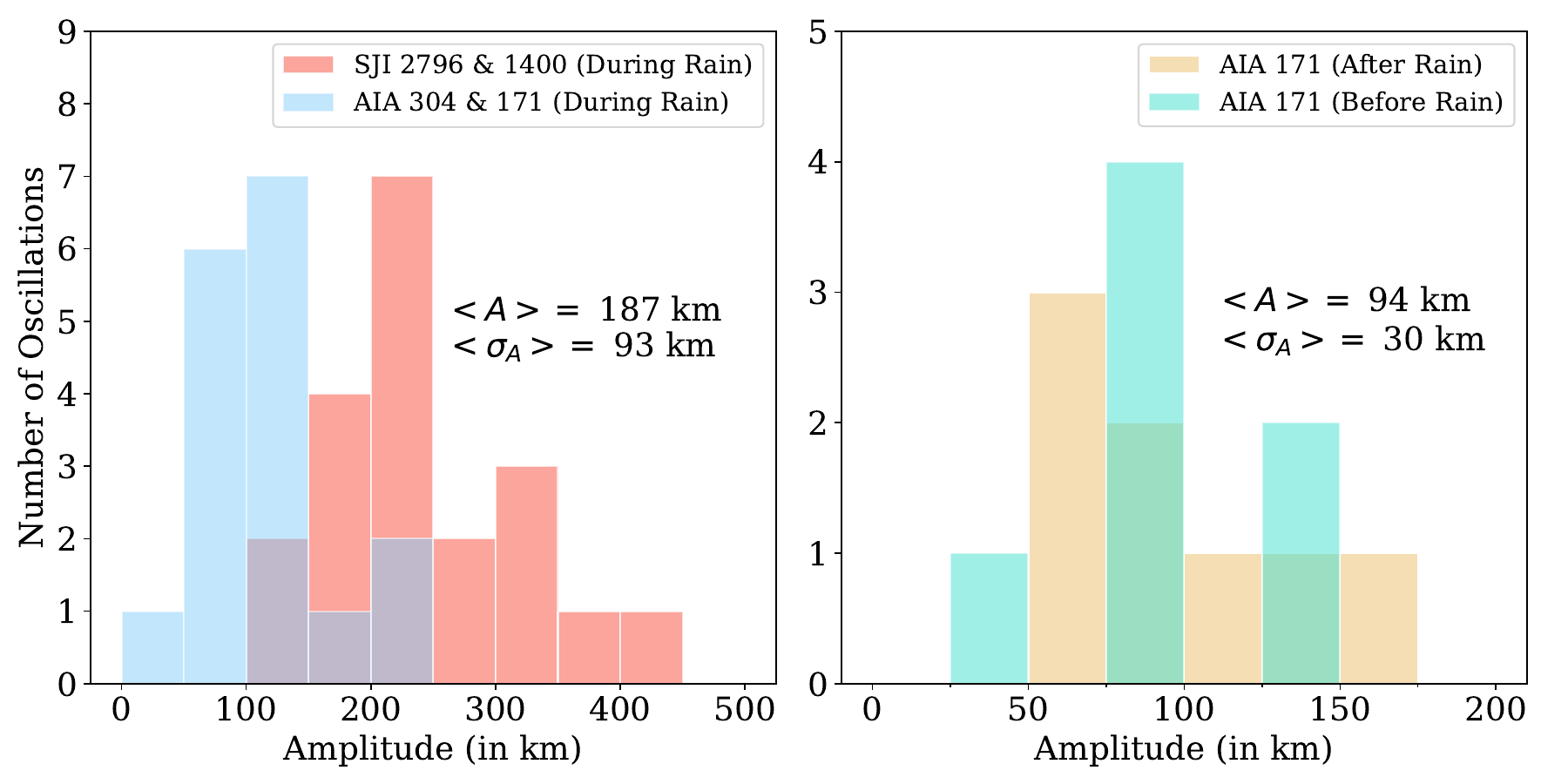}
              }
    \caption{The distribution of oscillation amplitudes captured during coronal rain (\textit{left panel}) and before and after coronal
rain (\textit{right panel}). $\left<A\right>$ and $\left<\sigma_{A}\right>$ denote average amplitude and standard deviation in amplitude, respectively.}
    \label{fig:all_ampl}
\end{figure}

\subsection{Comparison of average oscillations properties during, before, and after coronal rain}
\begin{figure}[!ht]
    \centering
    \centerline{
               \includegraphics[trim=0cm 0.3cm 0cm 0.3cm ,width=0.5\textwidth,clip]{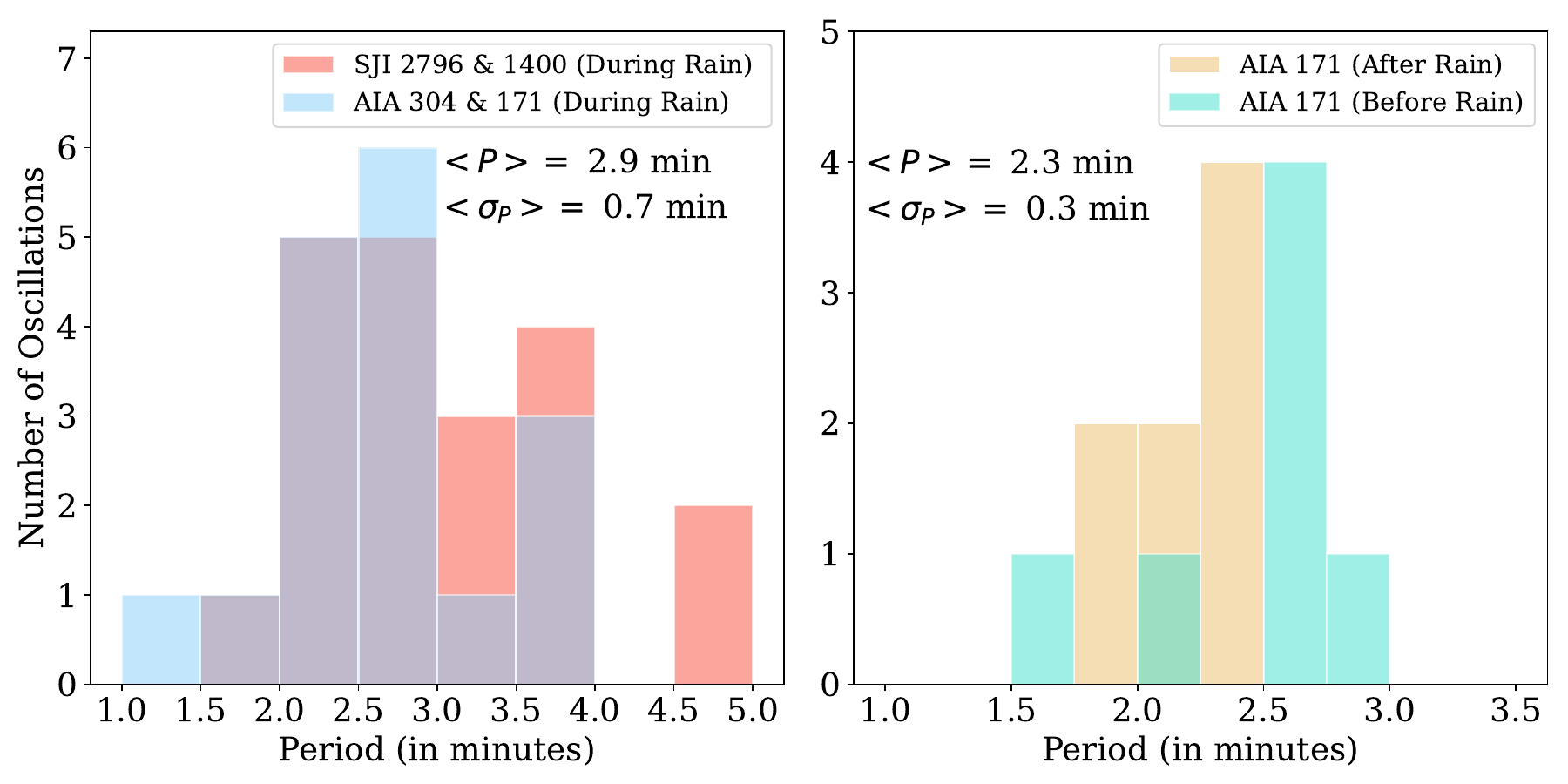}
               
              }
    \caption{The distribution of oscillation periods captured during coronal rain (\textit{left panel}) and before and after coronal
rain (\textit{right panel}). $\left<P\right>$ and $\left<\sigma_{P}\right>$ denote the average period and standard deviation in the period, respectively.}
    \label{fig:all_peri}
\end{figure}
We discussed individual cases of transverse oscillations captured before and after coronal rain in bigger and smaller loops at slit 3 and slit 8. Now, we consider the average properties of decayless oscillations captured in AIA 171 before and after coronal rain. Coronal rain is clearly observable in SJI 1400, 2796, and AIA 304, and oscillations during rain in both absorption and emission are identified in AIA 171. Consequently, we study the average properties of oscillations during coronal rain using these passbands. The left panel of Figure \ref{fig:all_ampl} shows the amplitude of oscillations captured by SJI 1400, 2796, and AIA 304 and 171 during rain. We find oscillations captured in these channels have amplitudes ranging from 43 to 415 km with an average of 187 km. Previous studies indicated amplitudes during coronal rain to be below 500 km(\citealt{antolin&verwitche2011}) and in the range of 200-400 km (\citealt{Kohutova&Verwichte2016}). The right panel of Figure \ref{fig:all_ampl} shows the distribution of oscillation amplitudes detected in AIA 171 before and after coronal rain in every slit position at similar locations as observed during rain. The amplitudes have a range of 45 to 164 km with a mean of 94 km. Such low amplitudes are earlier reported in various active-region loops (\citealt{2015Anfinogentov}, \citealt{Mandal}, \citealt{2022Zhong}). Thus, on average, oscillations during coronal rain have 2 times larger amplitudes than those before and after coronal rain. 

When we look at the distribution of period in Figure \ref{fig:all_peri}, it has a range of 1.4 to 4.6 minutes for oscillations detected during rain in SJI 1400, 2796, AIA 304 and 171 passband with a mean of 2.9 minutes, in agreement with earlier reported periods in the range of 1.4 to 3.3 minutes in coronal rain clumps \citep{antolin&verwitche2011}. In contrast, the periods range from 1.7 to 2.9 minutes, with a mean of 2.3 minutes for oscillations captured in the AIA 171 passband before and after rain. Thus, on average, oscillations during coronal rain have 1.3 times larger periods than those before and after coronal rain.
The average amplitude and period before coronal rain, calculated from AIA 171, are 93$\pm$28 km and 2.4 $\pm$0.4 minutes, respectively, while it is 94$\pm$34 km and 2.2$\pm$0.2 minutes after coronal rain.

We observe coronal rain oscillations in EUV absorption and possible signatures of the same in CCTR emission during coronal rain in AIA 171 x-t maps. The average amplitude of these oscillations is 90$\pm$31 km, and the average period is 2.5$\pm$0.2 minutes. The average amplitudes and periods are smaller compared to the average properties calculated from AIA 304 and SJI x-t maps. The oscillations in AIA 171 x-t maps during rain for slit 3 (O2 and O3 in Figure \ref{fig:fig5}) correspond to a time interval in which rain blobs started descending from the loop top to the footpoint. The oscillation period and amplitude can decrease during this interval, as observed previously in \cite{verwichte2017}. However, the periodic shift in the amplitude and period was not observed in AIA 304 x-t maps, as seen in the SJI 2796 x-t map of slit 1 (Figure \ref{fig:decay}). This could be due to the lower resolution of AIA compared to SJI.  The underestimation of displacement amplitudes compared to its actual value could be up to 20\% after motion magnification \citep{Gao}. This could also be a plausible explanation for the lower amplitude in AIA 304 x-t maps during coronal rain found in slit 3. To quantitatively confirm this, we reduced the resolution of SJI 2796 images to match AIA and performed motion magnification. After magnification, we analysed the oscillation in slit 3 and obtained a period identical to AIA 304. However, the amplitude is estimated to be 0.14$\pm$0.02 Mm for SJI 2796 oscillation, while it is 0.16$\pm$0.01 Mm for AIA 304, as discussed in section \ref{big_loop}. The amplitudes, considering their error bars, are quite similar.

\subsection{Calculation of loop geometry}
\begin{figure}[!ht]
    \centering
\includegraphics[trim=1.6cm 0cm 2cm 0cm ,width=0.5\textwidth,clip]{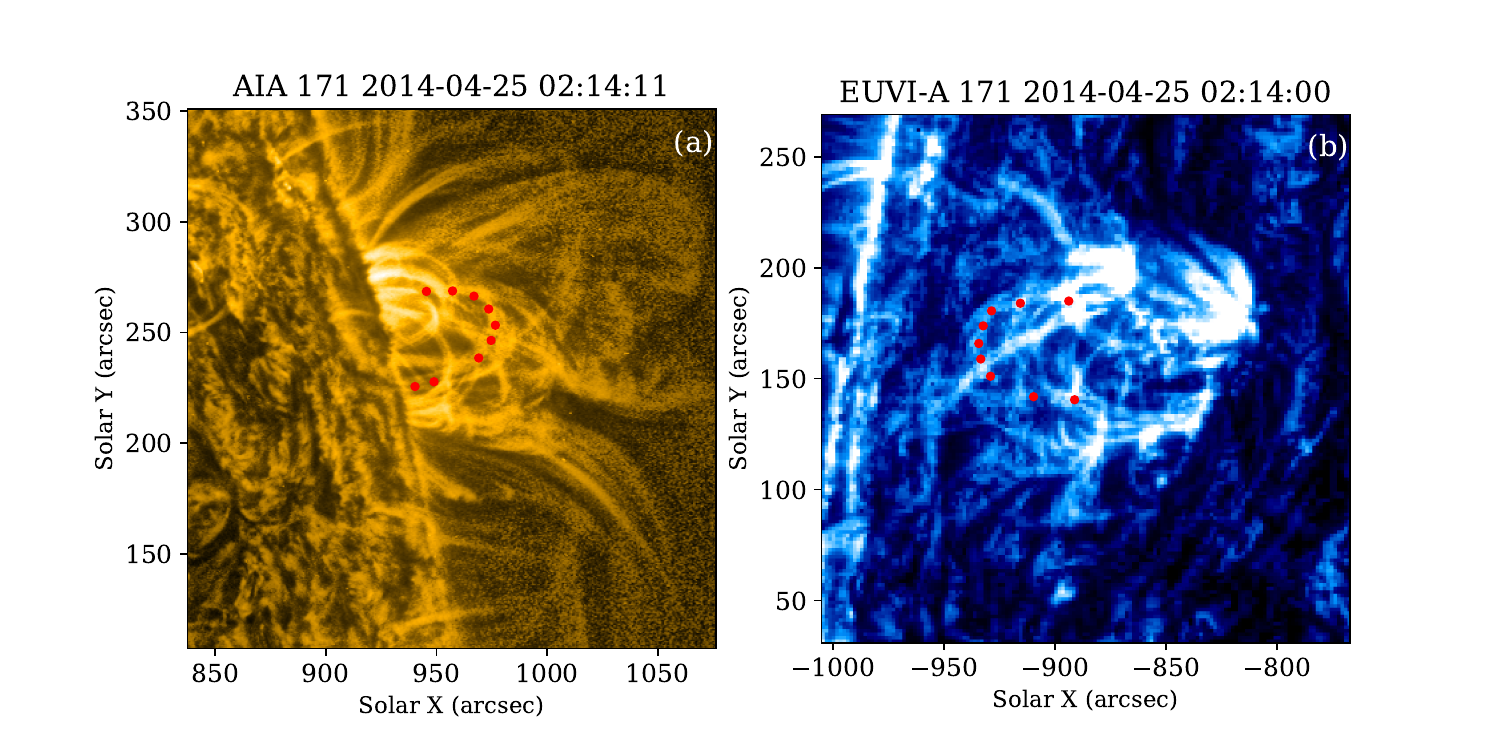}
\includegraphics[trim=0cm 0cm 0cm 0cm ,width=0.5\textwidth,clip]{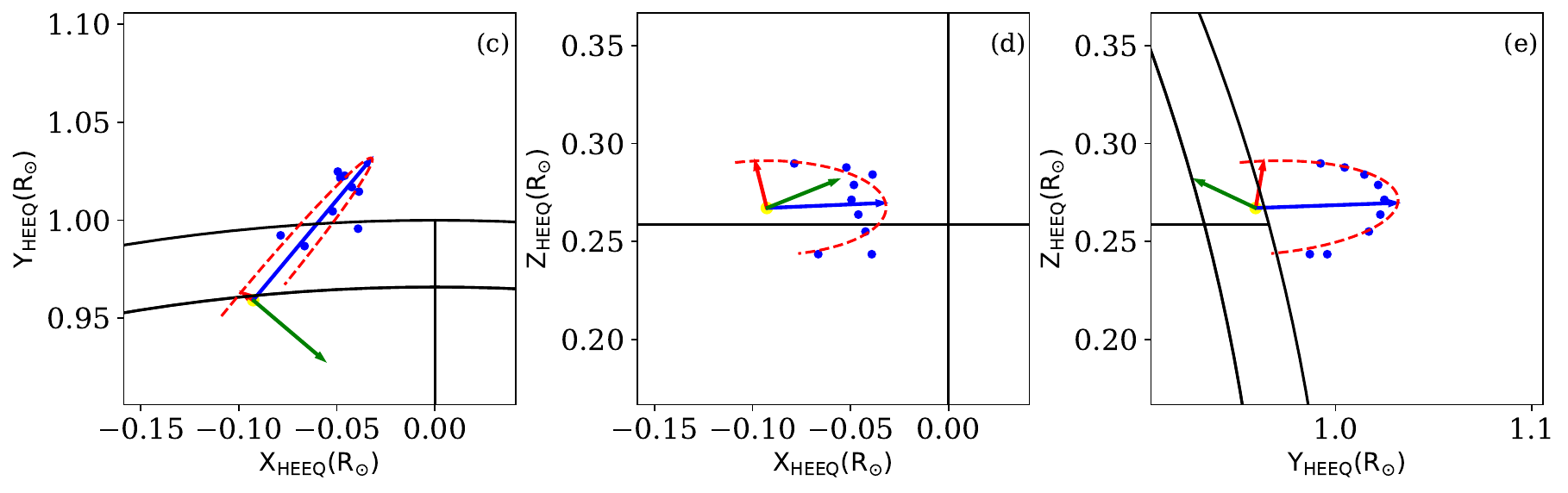}
    \caption{Panels (a)-(b) shows the region of interests as viewed by AIA 171 and EUVI-A 171. The images are processed using the MGN filter. The red points lie on the epipolar line and are used for 3D reconstruction. Panels (c)-(e) present the projections of the reconstructed loop in three different planes in the HEEQ coordinate system. The red dashed line indicates the fitted loop. The red and blue arrows show the minor and major axis of the fitted loop, while the green arrow indicates the normal to the loop plane.}
    \label{fig:loop_geometry}
\end{figure}

We estimate the 3D geometry of the bigger loop where slits 1 to 7 are positioned, and it is clearly visible in AIA images. We utilize the data from EUVI-A/STEREO located at 156.5\textdegree\ Stonyhurst longitude. The lower resolution of EUVI compared to AIA makes it difficult to identify the specific loop in EUVI-A images.  However, the similarity between structures seen in AIA and EUVI-A is exploited to distinguish the loop in EUVI-A. Figure \ref{fig:loop_geometry}(a)-(b) shows the region observed from two viewpoints (e.g. AIA and EUVI-A). We apply the MGN filter in both images to distinguish between loops better. We perform the stereoscopic triangulation using a python-based GUI tool \citep{2023Nistico}. The loop fitting is performed using Principal Component Analysis (PCA) \citep{2013Nistico, 2023Nistico}. The red points in Figure \ref{fig:loop_geometry}(a)-(b) represent the tie-points used for 3D reconstruction.

Figure \ref{fig:loop_geometry}(c)-(e) presents the projection of the 3d points in blue, calculated after triangulation, in the different orientations of the Heliocentric Earth Equatorial (HEEQ) coordinates system. The fitted loop is overplotted in red dashed curves. The loop is elliptical in shape with major and minor radii of 0.047 and 0.025 R$_\odot$, respectively. The loop length is estimated to be 140 Mm. The PCA fitting also provides normal to the loop plane, which can be used to estimate the inclination angle of the loop \citep{2013Nistico}. The inclination angle of the loop is found to be 36\textdegree\, which is a crucial parameter to obtain rain mass. We consider an error of 10\% in loop length and 10\textdegree\ in inclination angle.

\section{Discussion and conclusions} \label{sec:disc}
In this paper, we explore the effect of coronal rain on already present transverse oscillations in coronal loops observed by the AIA and SJI on April 25, 2014. The analysis of oscillations using SJI suggests oscillation amplitudes below 1 Mm during coronal rain, motivating us to use the motion magnification technique to capture oscillations in the AIA passbands. We take eight artificial slits in two coronal loops to generate x-t maps for before, during, and after coronal rain. The individual cases presented in this study show the signature of transverse oscillation at the same spatial positions before and after coronal rain in AIA 171. The individual cases show an increase in period and amplitude during coronal rain. The average amplitude during coronal rain is found to be 2.3 times larger than before and after coronal rain. The average period is found to be 1.3 times larger during coronal rain.

\cite{verwichte2017} confirmed the presence of coronal rain oscillation as soon as rain appeared in SJI channels. They also pointed out the presence of oscillations in an overlying loop in AIA 171 and 193 passbands. The periods in the overlying loop were similar, but amplitudes were larger than those found during coronal rain in the lower loop. The overlying loop was oscillating about 10 arcsec (7 Mm) above the lower loop. The oscillations were not present after coronal rain in AIA 171 or 193 (see figure 7 of \citealt{verwichte2017}). However, we see about a 1 Mm spatial difference in oscillations before, during, and after coronal rain in a few slits, which is only 8 Mm in length. The motion magnification allows us to quantify the oscillations having amplitude lower than the resolution of AIA, which was not the case for \cite{verwichte2017}.

We observe a larger mean period during coronal rain. The loop will be denser during coronal rain than prior and will be largely evacuated after coronal rain. In the long wavelength limit, kink speed will be equal to phase speed and inversely proportional to the period of oscillation (\citealt{Nakariakov2005}). The kink speed will decrease during coronal rain due to an increase in density, and eventually, the period of oscillation will be larger. Analytical studies and numerical simulations of transverse oscillations in a cooling coronal loop revealed that the period decreases in a cooling coronal loop with time (\citealt{Ruderman2011}, \citealt{Magyar2015}). The plasma cooling changes the density variation along the loop and produces flows from the loop top to the footpoints. \cite{Kohutava&Verwitche2017} pointed out using numerical simulation that as the condensations reach the loop legs from the loop top, the amplitude and period decrease and are converted into low amplitude, short period transverse oscillations. The decrease in the amplitude and period, seen after coronal rain, can be due to the rearrangement of loop plasma on a small time scale in the whole loop.

We can gain information on density changes due to rain using period variation. For kink oscillations, the square of the period will be proportional to the internal loop density for the constant magnetic field and outside density. Taking a crude assumption that the variation in density is simultaneous at each position of the loop \citep{verwichte2017}, we can write the following relation,

\begin{equation}\label{period_change}
   \rho_{d}/\rho_{b} = (P_{d}/P_{b})^{2} \; \; \&  \; \; \rho_{a}/\rho_{d} = (P_{a}/P_{d})^{2}
\end{equation}
where $\rho$ and P denote the internal density and period. The subscripts a, b, and d denote the after, before, and during coronal rain scenarios.
If we consider the period for oscillations detected in slit 3, which is near the loop top on the bigger loop (Figure \ref{fig:fig1}), then $\rho_{d}/\rho_{b} = \text{2.25}$ which means 125\% increase in the internal density in the formation of coronal rain. However, it should be noted that this formula is applicable for density variation occurring in the whole loop, while in coronal rain, there is a redistribution of mass and an increase in the density near the loop apex. Now, $\rho_{a}/\rho_{d} = \text{0.38}$. If we consider a constant volume of the loop, then approximately 60\% of rain mass is drained from the loop due to falling coronal rain.

Assuming that the observed amplitude during coronal rain is solely excited by coronal rain, we can calculate the rain mass fraction present during coronal rain using equation 39 of \cite{Verwitcheantoline2017}. Using oscillation parameters for oscillation detected in slit 3, loop length and inclination angle obtained after 3d reconstruction, we find the value of rain mass fraction to be 0.41$_{-0.14}^{+0.17}$. We observe that no coronal rain is present in the loop when oscillations are observed before and after coronal rain in AIA 171 (see Figure \ref{fig:slit_along_rain}). This indicates that the oscillations seen in AIA 171 before and after rain are not excited by coronal rain. It is suggested in numerical models and simulations that decayless oscillations can be triggered by footpoint drivers \citep{2016Nakariakov,2017Karampelas,2019Karampelas,2020Afanasyev}. Considering the footpoint driver being present all the time, the oscillations observed before and after coronal rain might result from footpoint driving. During rain, the observed oscillation properties could be influenced by two drivers: coronal rain and footpoint driving. The amplitude of oscillations can either increase or decrease due to coronal rain, depending on whether the two drivers are in phase or out of phase.

The amplitude during the coronal rain is larger than before, and given the small observed amplitudes of the oscillations, we can obtain the amplitude specifically associated with rain by subtracting the amplitude resulting from footpoint driving from the overall amplitude during the rain, taking into account the linear regime. In the context of slit 3 on the bigger loop, the amplitude attributable to footpoint driving is 0.08 Mm, detected from oscillation O1 in Figure \ref{fig:fig5}, while the amplitude during rain, considering the oscillation from SJI, amounts to 0.22 Mm. Consequently, the amplitude of oscillations solely attributed to rain is calculated as 0.14 Mm.

For the second loop (slit 8), the pre-rain amplitude is  0.09 Mm, obtained from oscillation O1 in Figure \ref{fig:fig4}. The initial amplitude during rain is 0.11 Mm, which grows by 2.1 times due to rain formation, resulting in the final amplitude being 0.23 Mm.  Assuming the amplitude before rain is a consequence of footpoint driving, the amplitude specifically due to rain is found to be 0.14 Mm, which is comparable to the amplitude of previously observed decayless oscillations \citep{2015Anfinogentov}.

We observe signatures of oscillations in AIA 171 during rain, potentially resulting from CCTR emission (Figure \ref{fig:fig5}). This implies that the amplitude of rain-driven oscillations can indeed be detected using AIA 171 alone, owing to the significant intensity produced by CCTR. If the analysis relies solely on AIA 171, there is a risk of incorrectly using the observed rain-driven oscillations in MHD seismology. The assumption of a coronal density that is lower than the actual value during rain would lead to an inaccurate estimation of local conditions in the loop. Assuming a rain density of 10$^{10}$ cm$^{-3}$, the calculated magnetic field value from MHD seismology will be $\sqrt{10}$ times larger compared to the value obtained using a coronal density of 10$^{9}$ cm$^{-3}$, with the density ratio considered to be zero. 

The results shown in this paper provide evidence of decayless oscillation before and after coronal rain at similar spatial positions. We conclude that increased density due to coronal rain formation can alter the period of already present transverse oscillations in coronal loops. Our study indicates that the contribution to the observed oscillation properties during rain can be attributed to coronal rain and footpoint driving.  To check the broad applicability of our results, we need more studies of coronal rain with transverse oscillations and numerical simulations of the same.

\begin{table*}[!ht]
    \centering
    \caption{Details of slits and parameters of oscillations}
    \label{Tabel1}
    \begin{tabular}{cccccccc}
    \hline
    \vspace {0.1cm}
        \textbf{Channel} & \textbf{Slit} & \textbf{Start Time (UT)} & \textbf{End Time (UT)} & \textbf{A$_{1}$ (km)}  & \textbf{P (mins)} & \textbf{$\tau$  (mins)} & \textit{\textbf{k}}  
        \vspace {0.1cm}
        \\ \hline

        SJI 1400 & S1 & 2:49:05 & 2:55:37 & 328 ± 14 & 2.9 ± 0.05 & - & - \\ 
        ~ & S1 & 3:02:47 & 3:09:39 & 237 ± 9 & 3.6 ± 0.04 & - & - \\ 
        ~ & S1 & 3:03:44 & 3:09:01 & 297 ± 10 & 2.6 ± 0.02 & - & - \\ 
        ~ & S1 & 2:52:49 & 2:56:52 & 241 ± 8 & 3.2 ± 0.06 & - & - \\ 
        ~ & S1 & 3:19:00 & 3:23:40 & 143 ± 7 & 2.2 ± 0.03 & - & - \\ 
        ~ & S2 & 2:51:34 & 2:56:52 & 243 ± 5 & 2.7 ± 0.02 & - & - \\ 
        ~ & S2 & 3:09:57 & 3:14:38 & 196 ± 7 & 3.1 ± 0.04 & - & - \\ 
        ~ & S2 & 2:56:33 & 3:02:47 & 415 ± 10 & 3.6 ± 0.03 & - & - \\ 
        ~ & S3 & 2:52:49 & 2:58:26 & 219 ± 6 & 2.7 ± 0.02 & - & - \\ 
        ~ & S4 & 2:51:53 & 2:55:56 & 199 ± 6 & 2.8 ± 0.08 & - & - \\ 
        ~ & S4 & 2:56:52 & 3:00:37 & 315 ± 12 & 2.1 ± 0.04 & - & - \\ 
        ~ & S5 & 3:17:45 & 3:22:25 & 265 ± 9 & 3.6 ± 0.09 & - & - \\ 
        ~ & S5 & 3:08:05 & 3:12:27 & 323 ± 21 & 3.3 ± 0.14 & - & - \\ 
        ~ & S7 & 3:06:51 & 3:10:35 & 394 ± 14 & 2.2 ± 0.02 & - & - \\ 
        ~ & ~ & ~ & ~ & ~ & ~ & ~ & ~ \\ 
        SJI 2796 & S1 & 3:01:23 & 3:18:32 & 348 ± 19 & 6.1 ± 0.09 & -13.9 & -0.14 \\ 
        ~ & S2 & 3:02:19 & 3:11:40 & 211 ± 6 & 4.6 ± 0.03 & - & - \\
        ~ & S3 & 2:54:32 & 2:59:50 & 142 ± 4 & 2.3 ± 0.02 & - & - \\
        ~ & S3 & 3:05:26 & 3:16:02 & 223 ± 3 & 3.9 ± 0.01 & - & - \\ 
        ~ & S4 & 3:10:07 & 3:15:44 & 214 ± 6 & 3.0 ± 0.02 & - & - \\ 
        ~ & S8 & 3:02:19 & 3:11:03 & 111 ± 10 & 2.1 ± 0.01 & 11.6 & 0.02 \\ 
        ~ & ~ & ~ & ~ & ~ & ~ & ~ & ~ \\ 
        AIA 304 & S1 & 3:04:55 & 3:09:43 & 227 ± 4 & 2.7 ± 0.02 & - & - \\ 
        ~ & S2 & 3:05:43 & 3:11:43 & 152 ± 4 & 3.6 ± 0.03 & - & - \\ 
        ~ & S3 & 3:07:43 & 3:17:07 & 155 ± 5 & 3.8 ± 0.09 & - & - \\ 
        ~ & S4 & 3:09:43 & 3:15:43 & 138 ± 3 & 3.6 ± 0.03 & - & - \\ 
        ~ & S5 & 3:02:43 & 3:07:19 & 140 ± 8 & 1.8 ± 0.02 & - & - \\ 
        ~ & S5 & 3:13:43 & 3:18:07 & 114 ± 3 & 3 ± 0.05 & - & - \\ 
        ~ & S6 & 3:05:43 & 3:09:07 & 212 ± 6 & 2.8 ± 0.05 & - & - \\ 
        ~ & S6 & 3:15:07 & 3:18:43 & 107 ± 3 & 2.5 ± 0.04 & - & - \\ 
        ~ & ~ & ~ & ~ & ~ & ~ & ~ & ~ \\ 
        AIA 171 & S1 & 2:28:23 & 2:35:59 & 85 ± 3 & 2.7 ± 0.01 & - & - \\ 
        ~ & S2 & 2:30:23 & 2:34:47 & 131 ± 5 & 2.6 ± 0.02 & - & - \\ 
        ~ & S3 & 2:30:47 & 2:36:11 & 84 ± 3 & 2.6 ± 0.02 & - & - \\ 
        ~ & S3 & 3:16:11 & 3:22:23 & 135 ± 3 & 2.4 ± 0.01 & - & - \\ 
        ~ & S3 & 3:15:47 & 3:23:23 & 79 ± 2 & 2.9 ± 0.02 & - & - \\ 
        ~ & S3 & 3:27:47 & 3:34:47 & 88 ± 6 & 2.4 ± 0.02 & - & - \\ 
        ~ & S4 & 2:28:59 & 2:36:11 & 84 ± 2 & 2.9 ± 0.01 & - & - \\ 
        ~ & S4 & 3:17:23 & 3:23:23 & 121 ± 2 & 2.5 ± 0.01 & - & - \\ 
        ~ & S4 & 3:16:47 & 3:23:23 & 80 ± 2 & 2.5 ± 0.01 & - & - \\ 
        ~ & S4 & 3:27:47 & 3:32:23   & 66 ± 4 & 2.3 ± 0.01 & - & - \\ 
        ~ & S5 & 2:27:47 & 2:34:59 & 129 ± 2 & 2.6 ± 0.01 & - & - \\ 
        ~ & S5 & 3:16:47 & 3:23:23 & 87 ± 2 & 2.4 ± 0.01 & - & - \\ 
        ~ & S5 & 3:17:59 & 3:23:23 & 61 ± 2 & 2.5 ± 0.02 & - & - \\ 
        ~ & S6 & 3:15:47 & 3:22:59 & 63 ± 1 & 2.2 ± 0.01 & - & - \\ 
        ~ & S6 & 3:17:59 & 3:23:23 & 43 ± 1 & 2.4 ± 0.02 & - & - \\ 
        ~ & S7 & 3:33:23 & 3:40:11 & 125 ± 3 & 2.4 ± 0.01 & - & - \\ 
        ~ & S7 & 3:16:59 & 3:20:35 & 62 ± 1 & 1.9 ± 0.01 & - & - \\ 
        ~ & S8 & 2:38:23 & 2:43:35 & 45 ± 2 & 2 ± 0.01 & - & - \\ 
        ~ & S8 & 2:36:23 & 2:40:11 & 88 ± 8 & 1.7 ± 0.04 & - & - \\ 
        ~ & S8 & 3:15:35 & 3:21:35 & 85 ± 3 & 1.7 ± 0.01 & - & - \\ 
        ~ & S8 & 3:23:59 & 3:28:35 & 72 ± 8 & 1.9 ± 0.02 & - & - \\ \hline
    \end{tabular}
\end{table*}

\begin{acknowledgements} \label{sec:ackn}

The authors express their gratitude to the anonymous referee for providing constructive feedback that contributed to the improvement of the paper. A.K.S is supported by funds of the Council of Scientific \& Industrial Research (CSIR), India, under file no. 09/079(2872)/2021-EMR-I. V.P. is supported by SERB start-up research grant (File no. SRG/2022/001687). P.A. acknowledges funding from his STFC Ernest Rutherford Fellowship (No. ST/R004285/2). V.P. acknowledges Dr. Erwin Verwichte for useful discussion.  A.K.S. and V.P. express gratitude to Tom Van Doorsselaere for insightful discussions. A.K.S. also acknowledges  Krishna Prasad S.,  Ritesh Patel,  Bibhuti Kumar Jha, and Satabdwa Majumdar for engaging and helpful discussions. We acknowledge NASA and SDO team for providing AIA data products. SDO is a mission for NASA's Living With a Star program. IRIS is a NASA small explorer mission developed and operated by LMSAL with mission operations executed at NASA Ames Research Center and major contributions to downlink communications funded by ESA and the Norwegian Space Centre.
\end{acknowledgements}
\vspace{-0.6cm}
\bibliographystyle{aa}
\bibliography{corain_oscl}

\end{document}